# Piezoelectricity of Atomically Thin WSe2 via Laterally Excited Scanning Probe Microscopy

Ehsan Nasr Esfahani [1,*], Terrance Li [2,*,†], Bevin Huang [2], Xiaodong Xu [2,3], and Jiangyu Li [1,4,‡]

1. Department of Mechanical Engineering, University of Washington, Seattle, WA 98195, USA
2. Department of Physics, University of Washington, Seattle, WA 98195, USA
3. Department of Materials Science and Engineering, University of Washington, Seattle, WA 98195, USA
4. Shenzhen Key Laboratory of Nanobiomechanics, Shenzhen Institutes of Advanced Technology, Chinese Academy of Sciences, Shenzhen 518055, Guangdong, China

## Abstract

Lattices of odd-layered two-dimensional (2D) transition metal dichalcogenides (TMDs) such as WSe$_2$ are non- centrosymmetric, and thus could possess linear piezoelectricity that is attractive for applications such as nanoelectromechanical systems (NEMS) and nanogenerators. Measuring the electromechanical coupling of 2D TMDs, however, is rather challenging, since its $D_{3h}$ point group symmetry makes conventional piezoresponse force microscopy (PFM) inapplicable. Here we develop a lateral excitation scanning probe microscopy (SPM) technique that enables mapping of the piezoelectric response of atomically thin WSe$_2$ directly on a substrate with high spatial resolution. Planar electrodes are used to excite piezoelectric vibrations while imposing anisotropic in-plane mechanical constraint to the WSe$_2$, resulting in an out-of-plane deformation due to Poisson's effect that can be measured by an SPM probe. Using this technique, we show that WSe$_2$ monolayer and trilayers exhibit strong electromechanical response linear to the applied excitation biases that is distinct from the substrate, while WSe$_2$ bilayer show negligible electromechanical response as expected. The effective piezoelectric coefficient is estimated to be 5.2 pm/V from the measurement, consistent with theoretical predictions. This method can be conveniently applied to a wide range of 2D materials with similar symmetry.

---

[*] These authors contributed equally to the work.
[†] A student at Newport High School in Bellevue, WA, and interned at UW Department of Physics.
[‡] Author to whom the correspondence should be addressed; email: jjli@uw.edu.

Piezoelectricity is the linear coupling of an applied electric field with the mechanical deformation of a material that interconverts mechanical and electrical energies [1]. It requires the absence of an inversion center in the crystal lattice, and in low-dimensional systems, is attractive for applications in nanoelectromechanical systems (NEMS) and nanogenerators [2–5]. One such class of low-dimensional materials are the two-dimensional (2D) transition metal dichalcogenide (TMDs) monolayers (ML) with the structure $MX_2$ (M = Mo, W; X = S, Se) [6]. Indeed, density functional theory (DFT) calculations predict that a host of TMDs MLs are piezoelectric, with coupling coefficients higher than that of traditionally employed bulk wurtzite structures [6,7]. Measuring the piezoelectricity of 2D materials experimentally, however, is challenging, since TMD MLs belong to the point group $D_{3h}$, and thus can only respond mechanically to an in-plane electric field. Nevertheless, apparent piezoelectricity has been observed in $MoS_2$ MLs on a flexible PET substrate [8,9], and the piezoelectric coefficient measured in $MoS_2$ ML samples suspended over a micro-fabricated gap [10]. In the latter experiment, an electric field applied through the planar electrodes generates an out-of-plane force in $MoS_2$ due to anisotropic in-plane constraint imposed by the electrodes, which was then measured via scanning force microscopy (SPM) probe. However, it is challenging both to map the piezoelectric response of 2D samples with high spatial resolution using such point-wise method and to transfer samples over a pre-patterned gap. To overcome these difficulties in such force-based measurements, we developed a lateral excitation SPM technique that enables the mapping of the piezoelectric response of 2D TMDs directly on a substrate with high spatial resolution. From this, we measured the effective piezoelectric coefficient of 5pm/V for $WSe_2$ ML, consistent with theoretical predictions.

**Results and Discussion**

From symmetry, $MX_2$ MLs belong to the point group $D_{3h}$, and only possess three nonzero piezoelectric coefficients $d_{22}$, $d_{21}=-d_{22}$, and $d_{16}=-2d_{22}$ in Voigt notation [1], while $MX_2$ bilayers (BLs) are centrosymmetric and thus not piezoelectric. This suggests that $MX_2$ ML can only respond mechanically to an in-plane electric field $E_2$, but not to out-of-plane electric field $E_3$. Furthermore, an in-plane electric field $E_2$ will only induce plane strains $\varepsilon_2= d_{22}E_2$ and $\varepsilon_1=-\varepsilon_2$ without any out-of-plane deformation in a free-standing sample, making it difficult to measure the piezoelectricity of atomically thin TMDs. Piezoresponse force microscopy (PFM), which has been widely used for characterizing electromechanical coupling of low-dimensional materials, is no longer applicable, since it excites the sample using an out-of-plane electric field applied through a conductive AFM probe tip in contact with the sample [11–13], as schematically shown in **Fig. 1(a)**. To highlight this shortcoming, we carried out PFM measurements on a WSe$_2$ sample micromechanically exfoliated on 285 nm Si/SiO$_2$ substrate using Scotch tape [14]. The topography of this sample is mapped in **Fig. 1(b)**, from which ML, BL, as well as bulk WSe$_2$ can be identified from line profiles displayed in **Fig. S1(a)**). Simultaneous PFM amplitude mapping shown in **Fig. 1(c)**, however, reveals no distinct contrast between WSe$_2$ regions of different thicknesses as well as between the WSe$_2$ flake and the substrate, demonstrating the inability to measure any piezoelectric response from the WSe$_2$ ML using conventional PFM. The slight variances in response across the substrate and flake may arise from electrostatic interactions [15,16], and a recent PFM work on MoS$_2$ attributed the observed piezoresponse to flexoelectricity [10,17,18].

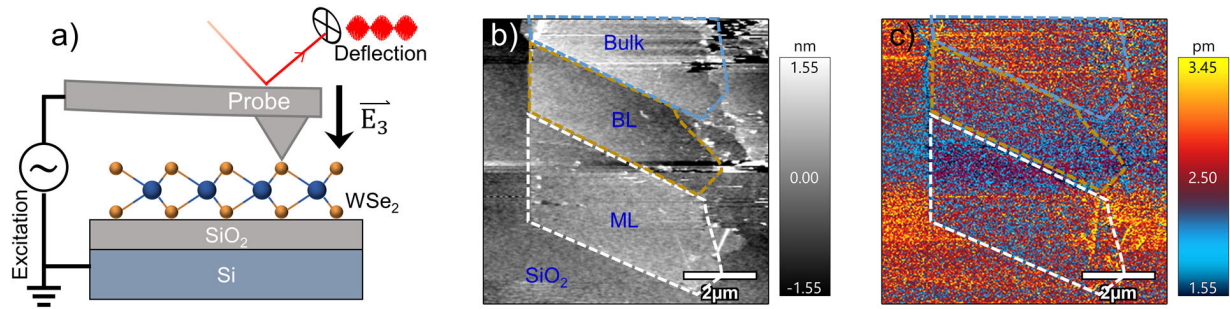

**Fig. 1** Conventional piezoresponse force microscopy (PFM) study of 2D WSe$_2$ on Si/SiO$_2$ substrate. (a) Schematic experimental setup; (b) Topography mapping revealing monolayer (ML) and bilayer (BL) WSe2; (c) PFM amplitude mapping showing no difference among ML and BL WSe$_2$ as well as the substrate.

Our adopted method is schematically shown in **Fig. 2(a)**, wherein two lateral electrodes are fabricated on both sides of atomically thin WSe$_2$ sample directly deposited on Si/SiO$_2$ substrate, enabling us to apply an AC bias to excite the piezoelectric vibration of the sample. Due to the anisotropic in-plane mechanical constraint induced by the electrode, an out-of-plane deformation is induced by the in-plane piezoelectric response. To appreciate this, we note that under the applied electric field $E_2$ and in-plane stress $\sigma_1$ and $\sigma_2$, the strains $\varepsilon_1$, $\varepsilon_2$, and $\varepsilon_3$ in the sample are given by

$$\epsilon_1 = -d_{22}E_2 + S_{11}\sigma_1 + S_{12}\sigma_2, \qquad \epsilon_2 = d_{22}E_2 + S_{11}\sigma_1 + S_{12}\sigma_2, \qquad \epsilon_3 = S_{13}(\sigma_1 + \sigma_2),$$

where in $S_{11}$, $S_{12}$, and $S_{13}$ are compliance constants. Clearly, in a free-standing WSe$_2$ without any external constraint, and thus in the absence of stress, an in-plane $E_2$ electric field will contract along its $x_1$-axis and expand along its $x_2$-axis by equal amounts, resulting in zero net vertical displace via Poisson's effect. However, out-of-plane strain can be induced if there is in-plane stress from the anisotropic constraint imposed by external electrodes. Assuming the extent of constraints are $\alpha_1$ and $\alpha_2$ along the $x_1$ and $x_2$ axes, respectively, then the stresses can be solved as

$$\sigma_1 = \frac{(1-\alpha_1)S_{11} + (1-\alpha_2)S_{12}}{S_{11}^2 - S_{12}^2} d_{22} E_2, \qquad \sigma_2 = \frac{(\alpha_2 - 1)S_{11} + (\alpha_1 - 1)S_{12}}{S_{11}^2 - S_{12}^2} d_{22} E_2,$$

which leads to an out-of-plane deformation $\varepsilon_3$ that is proportional to the applied electric field through piezoelectric coefficient $d_{22}$ via Poisson's effect,

$$\epsilon_3 = \frac{(\alpha_2 - \alpha_1) S_{13}}{S_{11} + S_{12}} d_{22} E_2 .$$

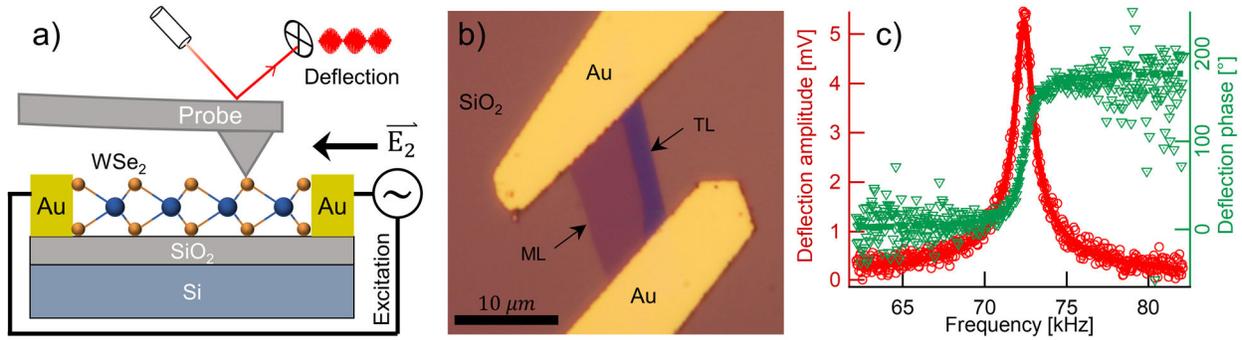

**Fig. 2** Lateral excitation SPM for measuring piezoelectric response of monolayer WSe$_2$. (a) Schematic experimental setup; (b) Optical microscope image of a WSe$_2$ device with gold electrodes; (c) measured amplitude versus excitation frequency.

This out-of-plane strain can be measured by the SPM probe, which is used here solely for sensing, unlike in a conventional PFM wherein the probe is used for both excitation and sensing. A non-conductive probe can then be used, minimizing possible interference by electrostatic interactions and issues with field concentration underneath the sharp tip. To demonstrate the concept, a WSe$_2$ device with lateral gold electrodes was fabricated, as shown in **Fig. 2(b)**, where a WSe$_2$ flake consists of both ML and trilayer (TL) between the two lateral electrodes are found, as detailed in **Fig. S1(b)**. By sweeping the AC bias frequency, a clear resonant peak in amplitude response is observed, as shown in **Fig. 2(c)**, across which phase flips by 180° as expected from a simple harmonic oscillator (SHO) model [19,20], confirming that the experimental setup is capable of exciting the electromechanical response of the 2D sample.

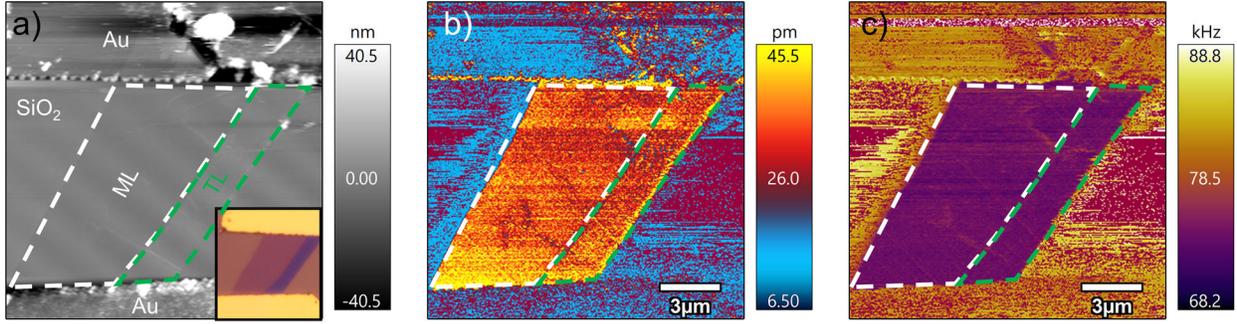

**Fig. 3** Mapping of electromechanical response of WSe$_2$ ML and TL. (a) Topography revealing WSe$_2$ ML and TL between gold electrodes. Inset: optical microscope image of the device; (b) Amplitude mapping; (c) Resonant frequency mapping.

We then move on to map the electromechanical response of WSe$_2$ on Si/SiO$_2$ substrate using lateral excitation SPM. **Fig. 3(a)** shows the topographical map of a representative WSe$_2$ ML and TL with its optical micrograph shown in the inset. Due to the different contact stiffnesses of the substrate and WSe$_2$ with the probe, the contact resonant frequency is expected to shift. To track this shift, we use a dual amplitude resonance tracking (DART) technique [21], with which the intrinsic amplitude and resonant frequency can be determined using an SHO model as well [20,22]. As seen in **Fig. 3(b)**, the amplitude mapping shows a clear distinction between the WSe$_2$ flake and the rest of substrate and electrode, in sharp contrast with conventional PFM in **Fig. 1(c)**, which is made evident in the direct comparison shown in **Fig. S2**. The WSe$_2$ sample exhibits an intrinsic piezoresponse of 35.3±5.8 pm (averaged across an 81 μm$^2$ area) under 5 V excitation bias with no differentiable difference between ML and TL, while the gold electrode and Si/SiO$_2$ substrate exhibit a much smaller and indistinguishable response of around 10.1±8.5 pm (spatial averaging over a 151 μm$^2$ area), which is believed to arise from an electrostatic background. By subtracting the substrate response, the effective piezoelectric coefficient of WSe$_2$ ML and TL can thus be estimated as 5pm/V, comparable to the theoretically predicted intrinsic piezoelectric coefficient of 5.1 pm/V [23]. Further, a large portion of substrate area (38.3 μm$^2$), highlighted by purple points, fails to yield a valid solution in the SHO model due to the low tip deflection amplitude. This contrast can also be seen from resonant frequency mapping in **Fig. 3(c)** showing lower stiffness of WSe$_2$ in comparison with the substrate. On the other hand, WSe$_2$ BL sample shown in **Fig. S1(c)** yielded tip deflection amplitudes similar to the SiO$_2$ substrate as shown in **Fig. S3**, wherein the substrate can hardly be distinguished from WSe$_2$. This further shows that the response from the atomically thin WSe$_2$ flakes is caused by the layer-dependent piezoelectricity between odd-numbered and even-numbered layers arising from (a lack of) centrosymmetry.

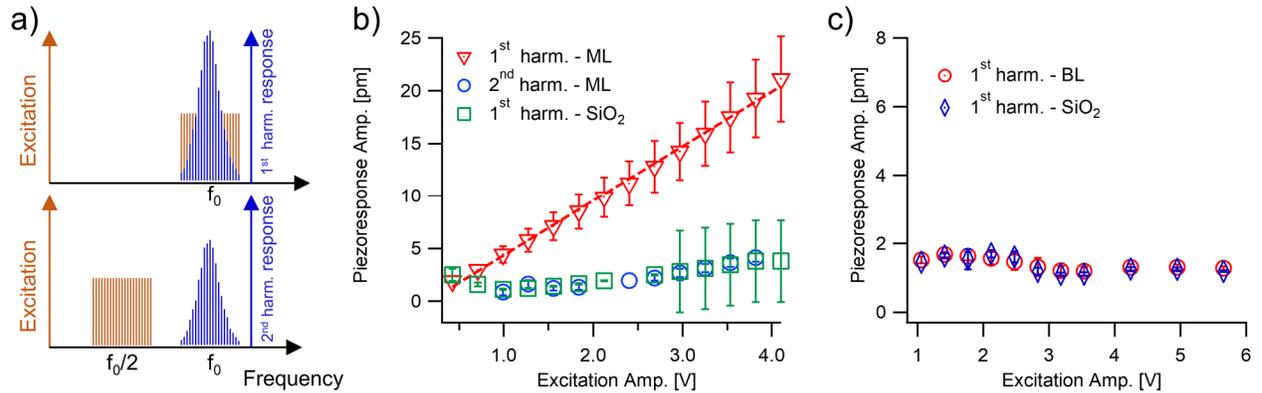

Fig. 4 Point-wise linear and quadratic electromechanical responses. (a) Schematic representation of first and second harmonic measurement for linear and quadratic response; (b) WSe$_2$ ML and TL showing dominant first harmonic response in sharp contrast with Si/SiO$_2$ substrate; (c) WSe$_2$ BL showing negligible first harmonic response indistinguishable from Si/SiO$_2$ substrate.

In order to confirm the linear piezoelectricity in atomically thin WSe$_2$, we carried out point-wise first and second harmonic measurements on both ML/TL and BL WSe$_2$, as schematically shown in **Fig. 4(a)**. A first harmonic experiment excites and measures the response at the fundamental contact resonance of $f_0$, and thus indicates a linear response of the tip deflection, while a second harmonic experiment excites at $f_0/2$ and measures at $f_0$, and thus measures the quadratic response of the tip deflection [15,24]. Such measurements have been carried out over a range of excitation biases up to ~4 V and on ten spatial points across the WSe$_2$ ML/TL flake and SiO$_2$ substrate (**Fig. 4(b)**), wherein it is clearly observed that the first harmonic response is linear with respect to the applied bias and dominates the second harmonic quadratic response. This shows that the electromechanical response from WSe$_2$ ML and TL results from linear piezoelectricity and not from ionic electrochemical dipoles induced by the charged probe [24]. The effective piezoelectric coefficient can be estimated from the slope, measured to be 5.2±0.08 pm/V, consistent with earlier values estimated from amplitude mapping in **Fig. 3(b)**. Furthermore, it is also noted that the first harmonic response of WSe$_2$ is much higher than that of Si/SiO$_2$ substrate. WSe$_2$ BL, on the other hand, shows much smaller first harmonic response that is indistinguishable from that of Si/SiO$_2$ substrate, as seen in **Fig. 4(c)**, consistent with its nonpiezoelectric nature. We also examined the first and second harmonic responses measured via conventional PFM for the ML WSe$_2$ sample in **Fig. 1**, as shown in **Fig. S4**. It is evident that first and second harmonic responses are comparable, and the WSe$_2$ ML and Si/SiO$_2$ are indistinguishable, confirming that the response measured under conventional PFM does not arise from linear piezoelectricity.

In summary, we have developed a convenient method to map the piezoelectric response of 2D materials quantitatively on their substrate, compatible with most device configurations. The method is based on lateral excitation and the Poisson effect, overcoming difficulty associated with conventional PFM for 2D materials. There is no need for micro-fabricated suspension, making the method widely applicable. Using this method, we measured the effective piezoelectric coefficient of WSe$_2$ ML to be 5.2 pm/V, consistent with theoretical predictions based on DFT.

**Materials and Methods**

*Device Fabrication*: To test the electromechanical properties of 2D samples, electrodes must be attached using electron beam lithography (EBL), metal evaporation, and wire bonding [25,26]. Solution of 6% polymethylmethacrylate (PMMA) in dichlorobenzene was spin-coated on the chip of interest as a photoresist under 1000 RPM for 10 seconds and 4000 RPM for 60 seconds. The spin-coated chips were baked at 180˚C for 3 minutes to cure the PMMA, after which they were patterned through electron beam lithography. The chip was then loaded into a metal evaporator and 5nm of vanadium (sticking layer) and 90nm of gold were evaporated onto the chip. After evaporation, the excess PMMA and V/Au were lifted off using a 30-minute dichloromethane bath, and the electrode pads were wire-bonded to larger electrodes.

*Conventional PFM*: Piezoresponse force microscopy (PFM) is performed using MFP-3D AFM (Asylum Research, Santa Barbara). Resonance-enhanced PFM images are obtained by measuring the amplitude and phase of probe deflection in contact mode with the excitation bias amplitude of 5V applied to the probe tip while the contact resonance frequency is tracked using the dual-AC resonance tracking (DART) technique. The intrinsic amplitude, frequency, phase, and quality factor mappings are calculated based on the SHO model. The contact-mode scan is obtained with the rate of 0.65 Hz/line using conductive probes with nominal in-air resonance frequency of 45 kHz (PPP-EFM, Nanosensors). The optical lever sensitivity and the spring constant of each probe is calibrated before the operation using a noncontact approach as described in Higgins *et al.* [27].

*Lateral Excitation PFM*: The experiments are identical to the conventional PFM except that the excitation bias (5 V) is rerouted to the gold electrodes on the two sides of $WSe_2$ sample for in-plane excitation. The amplitude and phase of probe deflection is measured in DART mode and the intrinsic mappings are similarly obtained. Experiments are performed using conductive probes with nominal in-air spring constants of 14 and 45 kHz (Arrow CONTPt, NanoWorld - PPP-EFM, Nanosensors).

*Point-wise Measurements:* The point-wise PFM measurements are obtained by measuring the frequency-dependent amplitude and phase response of probe deflection (probe transfer function as shown in **Fig. 1(c)**). In the first harmonic measurements (linear measurements) the excitation and detection frequencies are identical and are swept around the tip-sample resonance frequency, where the frequency-dependent amplitude and phase responses are recorded. In the second harmonic measurements the excitation frequencies are half of the detection frequencies. The amplitude and phase transfer functions are fitted to an SHO model where the intrinsic amplitude, frequency, phase and quality factor are found for a certain excitation amplitude. The process is repeated with several excitation bias amplitudes ranging from 0.7 to 5.7 V and over 10 random points on the $SiO_2$ substrate as well as the $WSe_2$ sample. The spatially averaged first and second harmonic intrinsic amplitude responses are illustrated as a function of excitation bias. The excitation bias is routed to the probe tip and to the Au electrodes in conventional and lateral PFM, respectively. The codes developed for these point-wise studies for Asylum Research AFMs are shared publicly and can be found in ref. [28].

**Supporting Information**

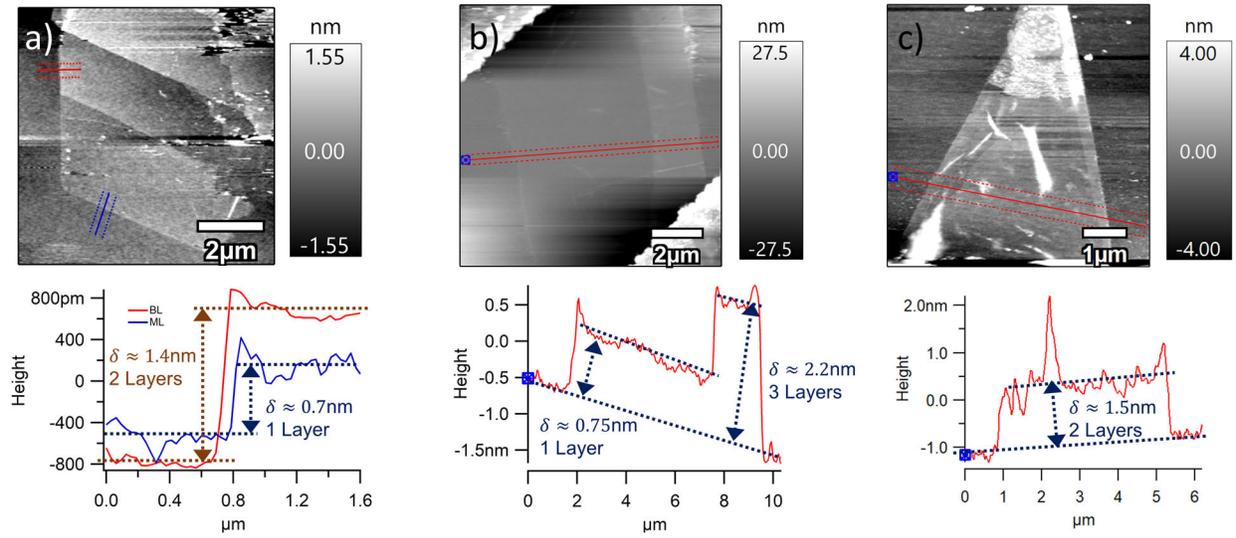

**Fig. S1** Topography mapping and line scans of three different WSe$_2$ samples.

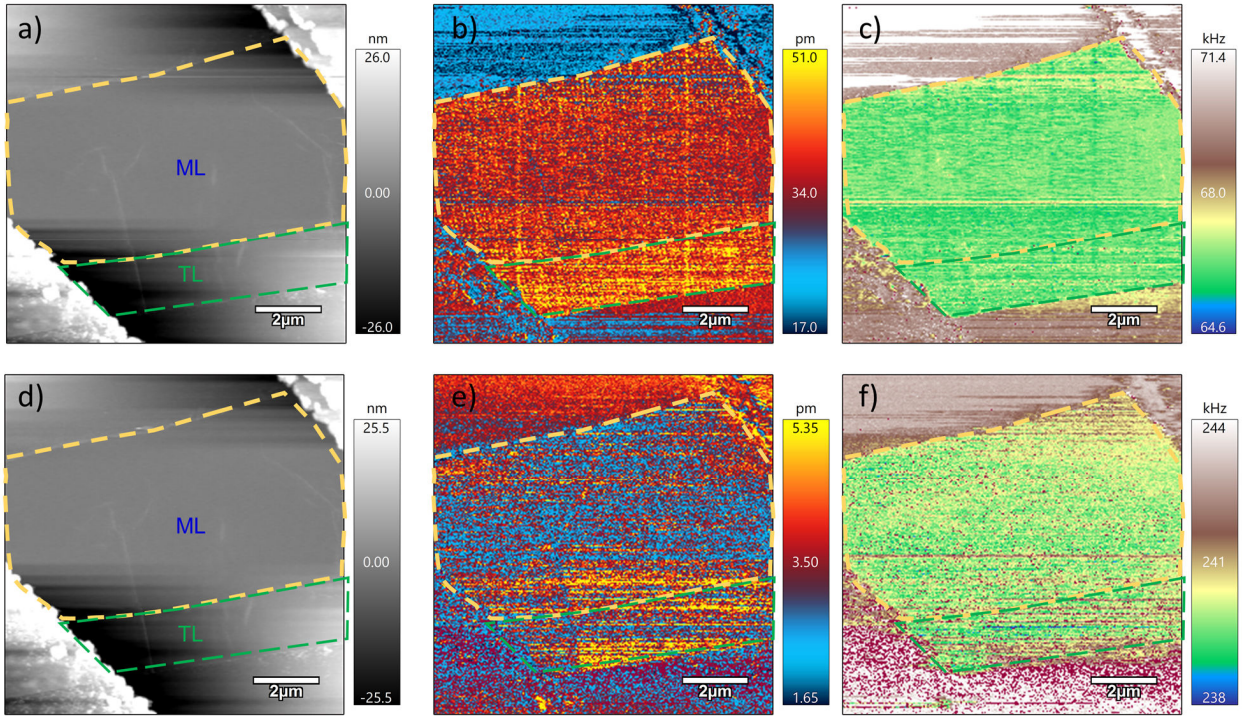

**Fig. S2** Comparison of our lateral excitation SPM (top row) and conventional PFM (bottom row) on monolayer WSe$_2$.

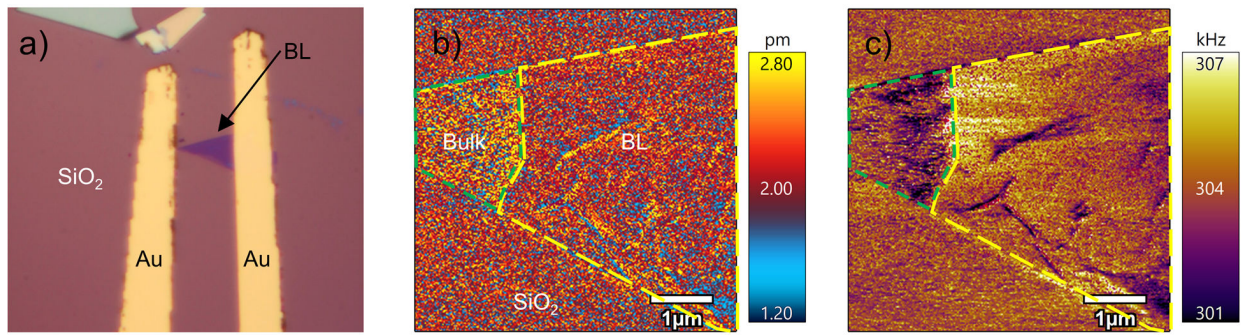

**Fig. S3** Lateral excitation PFM on BL WSe$_2$.

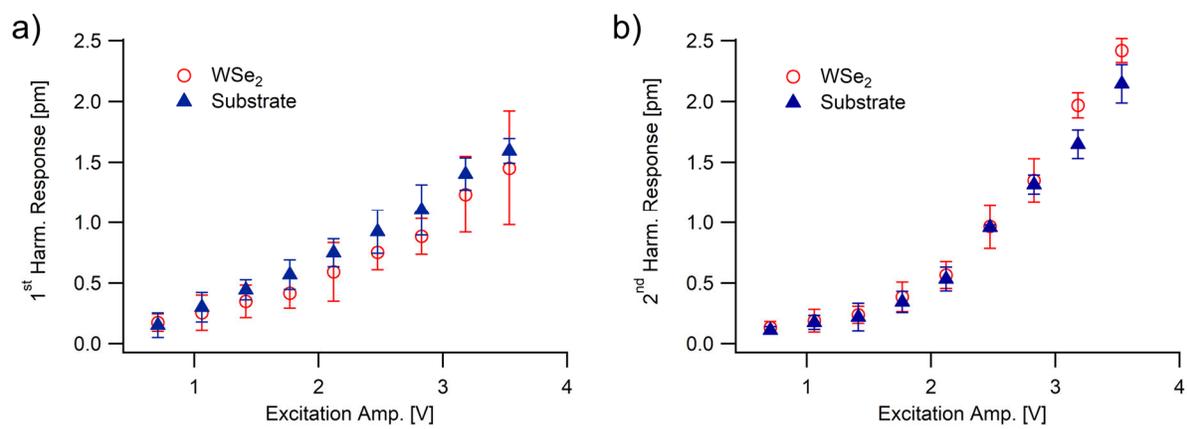

**Fig. S4** Conventional point-wise PFM study on a ML WSe$_2$ and the substrate.